%% file: arxiv_submission.tex
\documentclass[journal=apchd5,manuscript=article,layout=twocolumn]{achemso}
\usepackage[version=2]{mhchem} 
\usepackage{graphicx}
\usepackage{color}
\usepackage{verbatim}
\usepackage{float}
\usepackage{amsmath}

\author{Nathan Z. Zhao}
\affiliation[Stanford University]
{Department of Applied Physics, Stanford University, Stanford, California 94305, USA}

\author{Ian A. D. Williamson}
\affiliation[Stanford University]
{Department of Electrical Engineering, Stanford University, Stanford, California 94305, USA}

\author{Zhexin Zhao}
\affiliation[Stanford University]
{Department of Electrical Engineering, Stanford University, Stanford, California 94305, USA}

\author{Salim Boutami}
\affiliation[Stanford University]
{Department of Electrical Engineering, Stanford University, Stanford, California 94305, USA}
\alsoaffiliation[Grenoble]
{Universit\'{e} Grenoble Alpes, CEA, LETI, Grenoble, 38054, France}

\author{Shanhui Fan}
\affiliation[Stanford University]
{Department of Electrical Engineering, Stanford University, Stanford, California 94305, USA}
\email{shanhui@stanford.edu}

\date{today}

\title[An \textsf{achemso} demo]
  {Penetration depth reduction with plasmonic metafilms}

\begin{document}

\begin{abstract}

In many optical systems, including metal films, dielectric reflectors, and photonic crystals, electromagnetic waves can experience evanescent decay. The spatial length scale of such decay defines the penetration depth. The ability to reduce the penetration depth is important for a number of applications in free-space and integrated photonics. In this paper, we consider a ultrathin metafilm structures consisting of alternating regions of metal and dielectric. We show that the penetration depth of such metafilm can be significantly smaller as compared to that of a corresponding metal thin film. The reduction of the penetration depth arises due to the enhanced effective mass in the photonic band structure. This effect can be used to enhance the reflectivity of ultrathin reflectors, and to increase the packing density of subwavelength plasmonic waveguides. 

\end{abstract}

\maketitle

\section{Introduction}

In the optical wavelength range, due to the plasmonic response of metals, electromagnetic fields becomes evanescent inside of metals: at a given frequency, the amplitude of the fields decay exponentially as they penetrate into the metal. The length scale corresponding to such exponential decay defines the penetration depth. In typical plasmonic metals, the penetration depth can be far smaller as compared to the wavelength. Such a subwavelength penetration depth at optical frequencies is essential for many applications, including sub-wavelength metal films acting as mirrors \cite{Griffiths} and plasmonic waveguides which provide deep subwavelength modal confinement \cite{Zia2004, Oulton2008, Yan2011, Fang2015, Li2017}. The ability to design structures with penetration depth below that of natural metals would lead to dramatically improved device performance, particularly in the optical regime where absorption losses can be very high.



\begin{figure*}
    \centering
    \includegraphics[width = 5.9 in]{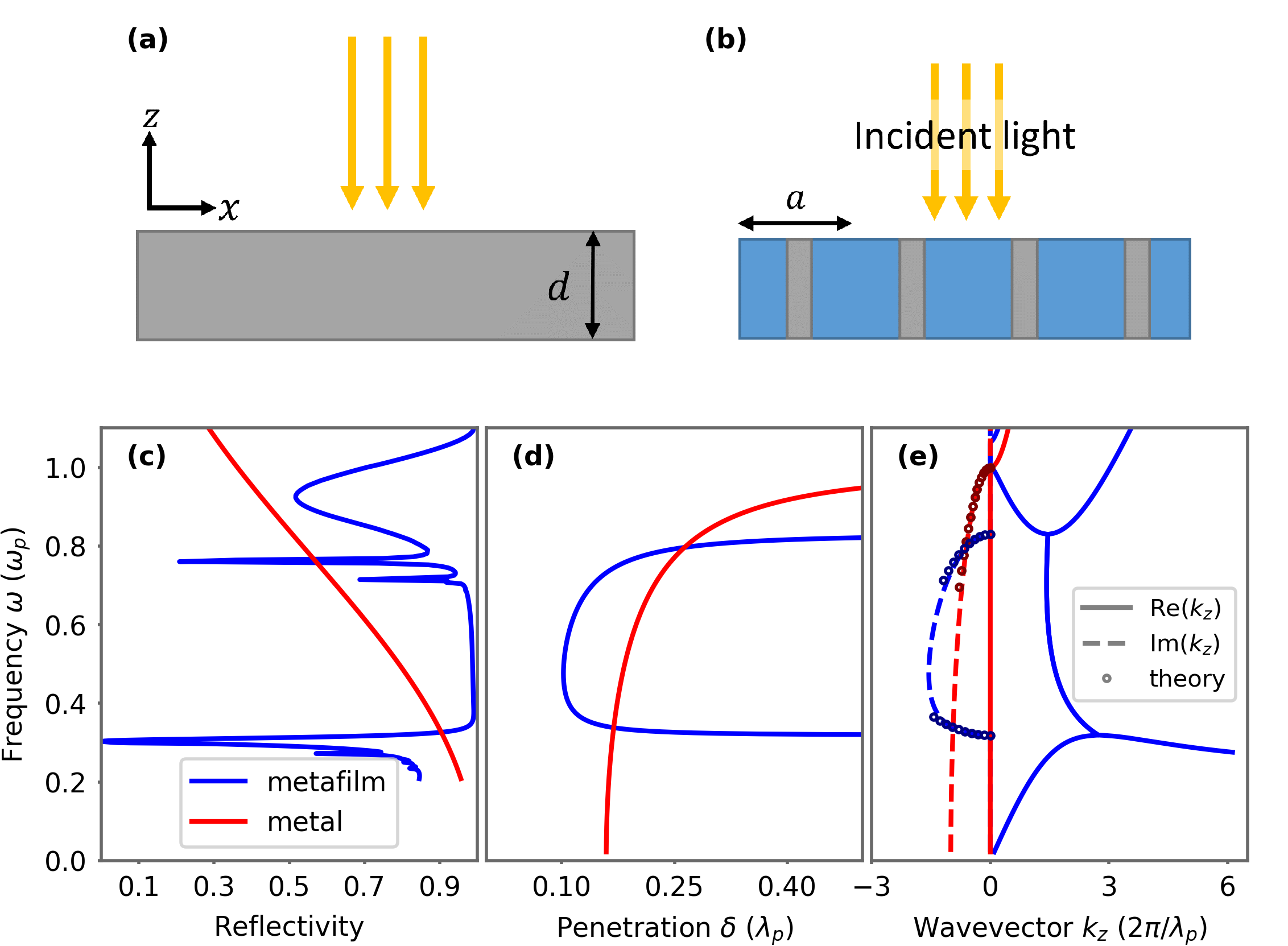}
    \caption{Diagram of (a) uniform metal and (b) metafilm reflector illuminated by normally incident light. The metal material (grey region) in both structures is described by a Drude model with a dielectric constant $\epsilon_m = 1-\omega^2/\omega_p^2$. The dielectric material (blue region) in the metafilm is described by a permittivity $\epsilon_d$. (c) Reflectivity spectrum for the metal (red) and metafilm (blue) structures with thickness $d = 0.24 \lambda_p$. The metafilm has a lattice constant $a = 0.24\lambda_p$ and $\epsilon_d = 16$. (d) Penetration depth and (e) real band structure (solid lines) and imaginary band structure (dashed lines) of the infinitely-extended structures, which correspond to finite versions of those in panel c. The analytic continuation, computed using a second-order approximation of the real band near the band-edges, is indicated by circles.}
    \label{fig:intro}
\end{figure*}

In this paper, we show that one can reduce the penetration depth of a one-dimensional plasmonic metafilm below that of standard metals using concepts from metamaterial engineering. As an illustration, we compare a uniform metal film, as shown in Fig. \ref{fig:intro}(a), to a metal-dielectric metafilm, as shown in Fig. \ref{fig:intro}(b). The metafilm consists of a periodic array of metal and dielectric regions. Both the periodicity and the thickness of the film are substantially smaller than the operating optical wavelength. Using full-wave numerical modeling, we show that for light incident on such a metafilm from air, the penetration depth can be far smaller as compared to the uniform metal thin film. Consequently, the metafilm can exhibit a substantially higher reflectivity and lower absorption as compared to a metal film with the same thickness. We also demonstrate that replacing the sidewalls of a metal-insulator-metal (MIM) waveguide with the metafilm results in an improved modal confinement and a reduced modal attenuation. The operation of our proposed metafilm relies upon the existence of a modal gap between the plasma frequency of the metal and the surface plasmon frequency of the metal-dielectric interface. We show that reduced penetration depth in this structure arises from the the enhanced effective mass of the photonic bands at the edges of the modal band gap. 

The remainder of this paper is organized as follows. First, we outline the theory for field penetration into the metafilm based on a band structure analysis of the corresponding infinitely-extended structure. Next, we illustrate the influence of the penetration depth on the reflection from a finite-thickness metafilm. We also identify features in the reflectivity spectrum, such as surface plasmon resonances, which are not accounted for in the band structure of the infinitely-extended structure. We then analyze the performance of the metafilm design constructed from several exemplary realistic materials and demonstrate that the effect is largely robust to intrinsic material absorption. Finally, we demonstrate an application of the metafilm as the sidewall of a compact hollow-core waveguide system.

\section{Band Structure Analysis}

For both the standard metal thin film and our metafilm structure, the evanescent behavior of the field penetration arises due to the existence of a photonic band gap in the corresponding structure which is infinitely extended in the $z$-direction. Such an infinite structure supports a band gap in the frequency range between $\omega_l$ and  $\omega_h$, where $\omega_l$ and $\omega_h$ are the lower and upper edges of the gap. For illustrative purposes, consider a frequency $\omega$ which is in the vicinity of the upper edge, $\omega_h$. For $\omega>\omega_h$, suppose the structure supports a band given by,
\begin{equation}
    \omega(k_z)=\omega_h + \frac{1}{2m_h} (k_z-k_h )^2,
    \label{eq:quadratic_dispersion}
\end{equation}
which describes a propagating wave in the $z$-direction. Here $m_h$ is the effective mass of the upper band, and $k_h$ is the wavevector at the upper band-edge. For $\omega<\omega_h$, the frequency is inside the gap. The field therefore varies exponentially along the $z$-direction in the form of $e^{-\text{Im}(k_z) z}$, where $\text{Im}(k_z)$, the spatial decay rate of the field amplitude, can be determined through the analytic continuation of the band structure of Eq. \eqref{eq:quadratic_dispersion} as,
\begin{equation}
    \text{Im}(k_z)=\sqrt{2m_h (\omega_h-\omega)}.\textbf{}
    \label{eq:kappa}
\end{equation}
$\text{Im}(k_z)$ is related to the penetration depth of the field amplitude as, $\delta = 1/\text{Im}(k_z)$. We therefore observe that the penetration depth inside the gap is directly related to the effective mass of the band structure in the frequency range just outside the gap. Such a relationship has been previously used to qualitatively describe anomalously high mode confinement in two-dimensional photonic crystals \cite{Ibanescu2006}. 

For the uniform, metallic thin film structure as shown in Fig. \ref{fig:intro}(a), the corresponding infinite structure is a bulk metal. To describe the bulk metal, we use the Drude model to describe its dielectric function $\epsilon_m(\omega) = 1-\frac{\omega_p^2}{\omega^2 + i \gamma \omega}$,  where $\gamma$ is the damping parameter or intrinsic loss. A direct solution of Maxwell's equation with $\gamma = 0$ gives the bulk band structure $\omega^2 = \omega_p^2 + c^2 k^2$. In the vicinity of the band-edge $\omega \approx \omega_p$, the band can be well described using Eq. \eqref{eq:kappa} with $\omega_h = \omega_p$ and $m_h = \omega_p/c^2$. The spatial decay rate $\text{Im}(k_z)$, as obtained using Eq. \eqref{eq:kappa} and shown in Fig. \ref{fig:intro}(e) as maroon circles, agrees well from the direct solution of the spatial decay rate using Maxwell's equations as shown by the red dashed lines in Fig. \ref{fig:intro}(e).


We now focus on describing the metafilm. The unit cell size is chosen to be subwavelength $a \ll \lambda$ with a fraction, $f$, consisting of a metal described by the permittivity $\epsilon_m$. The remaining fraction of the unit cell, $1-f$, consists of a high-index dielectric material with permittivity $\epsilon_d$. The thickness of the metafilm is $d$. As an exemplary structure, we choose $f = 0.2$, $\epsilon_d = 16$, and $a$ = $0.24 \lambda_p$.

For the metafilm structure, the corresponding infinite structure is a metal-dielectric waveguide array extended infinitely with the same dielectric and geometrical parameters as the film.  We calculate its band structure for waves propagating along the $z$-direction using a finite difference frequency domain eigensolver \cite{Veronis2005}, which solves Maxwell's equations in the $x$-$z$ plane for the TM polarization (non-zero vector field components $H_y$, $E_x$, $E_z$). In this solver, we obtain the wavevector $k_z$ for a corresponding real frequency, $\omega$. The computed band structure, $\omega{(k_z)}$, is plotted in Fig. \ref{fig:intro}(e). The frequency range where the wavevector is real is the range for the photonic bands, whereas the frequency range where the wavevector is complex is the photonic band gap region. For this structure, in the frequency range below $\omega_p$, we observe two bands separated by a band gap. Both the lower band-edge of the upper band, and the upper band-edge of the lower band, are located at band-edge wavevectors which are non-zero. Inside the gap, the real part of the wavevectors, Re($k_z$), connects between the band-edge wavevectors. The imaginary part of the wavevector, Im($k_z$), vanishes at the edges of the gap, and reaches its maximum slightly below the gap center. 

To understand the behavior of the imaginary part of the wavevector inside the gap, we again fit the upper band-edges using Eq. (\ref{eq:quadratic_dispersion}). Using the fitting parameters, we analytically determine $\text{Im}(k_z)$ using Eq. \eqref{eq:kappa}, and compare to the numerically calculated Im($k_z$) shown in Fig. \ref{fig:intro}(e). The analytic results agree very well with the numerical results. Similar agreement is obtained for the frequency range within the gap near the lower band-edge. Thus, we show that the field penetration behavior of this waveguide array structure inside the gap can be well accounted for by using analytic continuation of the photonic bands that describe the propagating waves.

Within much of the band gap, the waveguide array structure has an Im($k_z$) that is significantly larger as compared to that of the bulk metal. Thus, the field penetration depth of the waveguide array structure can be significantly smaller than that of the bulk metal, as shown in Fig. \ref{fig:intro}(d). This is remarkable given the fact that the waveguide array considered here is mostly made of dielectric and hence has far less metal as compared to the bulk metal structure. From the band structure analysis above, the reduction of the field penetration depth in the waveguide array structure is directly correlated with the enhanced effective mass at the band-edges as compared with the bulk metal structure, as can be seen by examining the band structure in Fig. \ref{fig:intro}(e).

\begin{figure}
    \centering
    \includegraphics[width = 3.25 in]{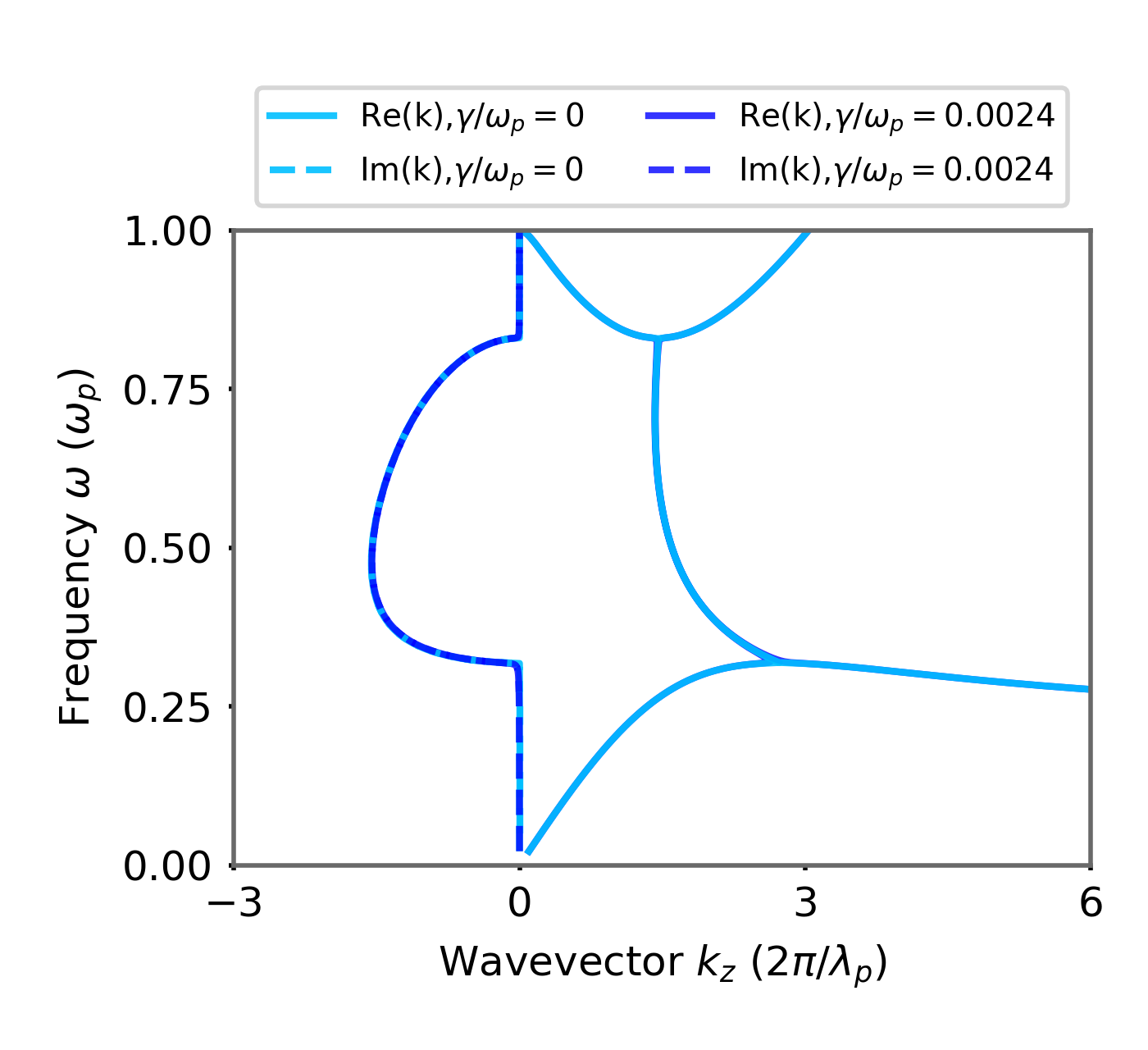}
    \caption{The real and imaginary band structure for the metafilm constructed from a dielectric of permittivity $\epsilon_d = 16$ and metal described by a lossless (light blue) and lossy (blue) Drude model. The lossy Drude model parameters correspond to those for silver.}
    \label{fig:loss_analysis}
\end{figure}

The final part of our band structure analysis is to consider the effect of material absorption. For this purpose we consider the metal as described by a lossy Drude model $\epsilon_m(\omega) = 1 - \omega_p^2/(\omega (\omega + i \gamma))$. Here we choose $\gamma = 0.0024 \omega_p$, which is appropriate for silver \cite{Blaber2009}. Comparing the lossless and lossy cases (Fig. \ref{fig:loss_analysis}), we see that the presence of the loss does not significantly change the band structure. And hence the field penetration behavior, as described by Im$(k_z)$, is not significantly affected by the presence of such loss. 


\section{Spectra Analysis and Effects of Finite Thickness}

Based on the analysis of the band structures of the two structure in the previous section, which are applicable only for structures which are infinite in the $z$-direction, we now consider the reflection from finite structures. We use a rigorous coupled wave analysis (RCWA) method to compute the reflectivity of a metafilm with a thickness $d=0.24 \lambda_p$. We use the lossy Drude model as described above for Fig. \ref{fig:loss_analysis}, where $\gamma = 0.0024 \omega_p$\cite{Blaber2009}. The red line in Fig. \ref{fig:intro}(c) corresponds to the reflectivity spectrum  of a metal film with the same thickness. The reflectivity increases as frequency drops further below the plasma frequency, $\omega_p$. The blue line in Fig. \ref{fig:intro}(c) corresponds to the reflectivity spectrum of the metafilm. Inside the band gap, the metafilm generally exhibits a much higher reflectivity compared with the uniform metal film structure, which is consistent with the prediction of a reduced penetration depth from the previous section. 



\begin{figure}[H]
    \centering
    \includegraphics[width = 3.25 in ]{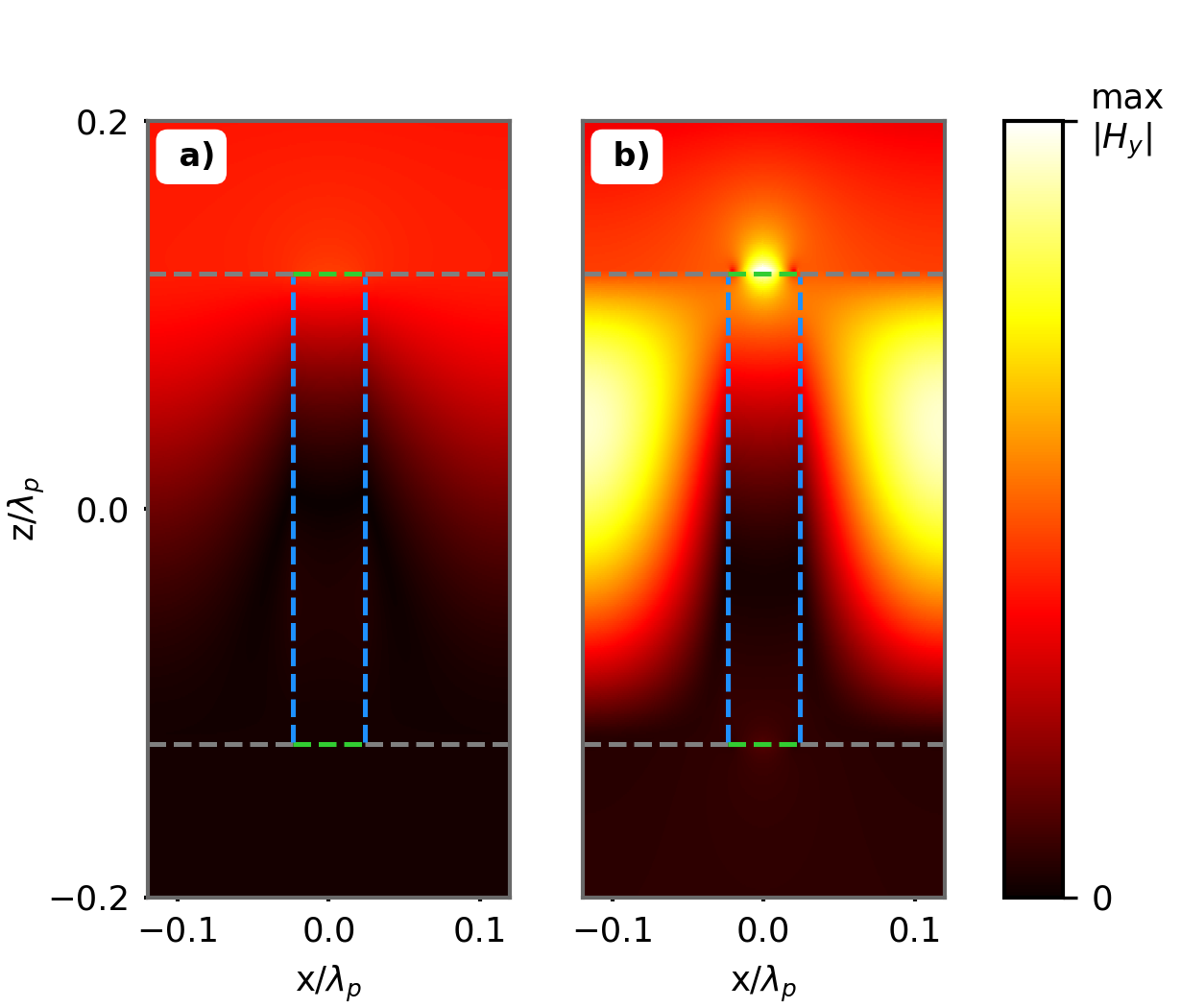}
    \caption{(a) Distribution of $\vert{H_y{(x,z)}}\vert$ when a plane wave with a frequency $\omega = 0.45 \omega_p$, which is near the frequency of minimum penetration depth, is normally incident upon the metafilm of thickness $d = 0.24 \lambda_p$. The dashed gray line denotes the dielectric-air interface, the dashed green line denotes the metal-air interface, and the dashed blue line denotes the metal-dielectric interface. (b) Distribution of $\vert{H_y{(x,z)}}\vert$ at a frequency of $\omega = 0.71\omega_p$, where the the surface plasmon resonance on the metal-air interface is clearly visible.}
    \label{fig:resonance_dip}
\end{figure}

Comparing the reflectivity spectrum of the metafilm [Fig. \ref{fig:intro}(c)] with the band structure of its corresponding infinitely extended structure, we note the lower frequency edge where the finite structure exhibits strong reflection agrees quite well with the lower band-edge of the infinite structure. The metafilm structure exhibits strong reflection over much of the frequency range of the gap. The field pattern for the frequency $\omega = 0.45\omega_p$, which lies near the region of minimal penetration depth, is shown in Fig. \ref{fig:resonance_dip}(a), where we indeed see the strong attenuation of the field inside the film.  Near the upper band-edge, in the band gap region, the finite structure exhibits two dips in the reflectivity spectrum. These dips correspond to localized surface plasmon modes at the metal-air interfaces, as shown in Fig. \ref{fig:resonance_dip}(b), where we plot the field at one of the resonance peaks located at $\omega = 0.71 \omega_p$, which is approximately the surface plasmon frequency. These resonances are highly absorptive.


\begin{figure}[H]
    \centering
    \includegraphics[width = 3.2 in, height = 4.7 in]{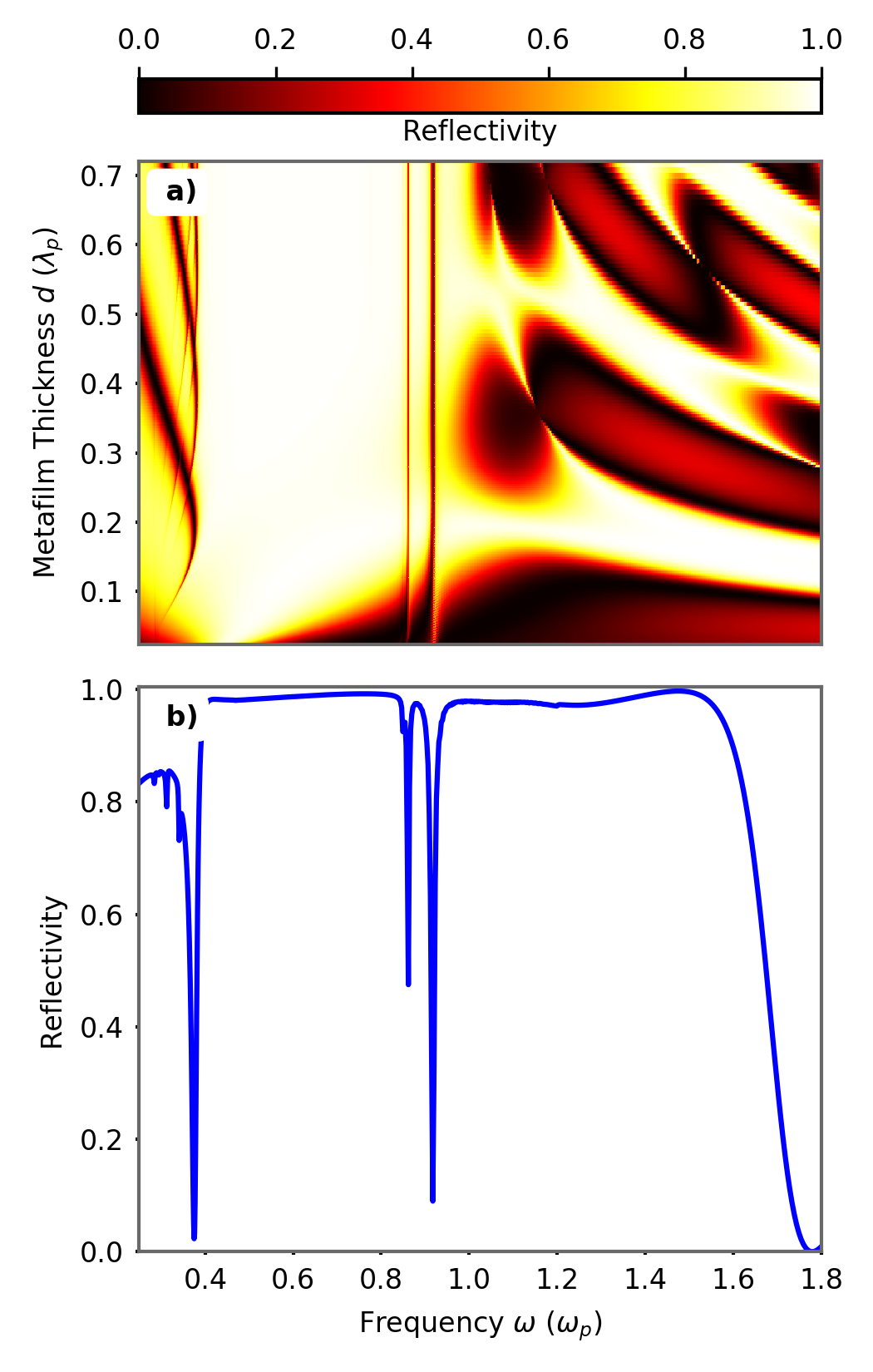}
    \caption{(a) The reflectivity as a function of the metafilm thickness $d$ and the frequency $\omega$. The structure has a lattice constant of $a = 0.24 \lambda_p$ and a metal filling fraction $f = 0.2$.  (b) Selected spectrum from (a) with a metafilm thickness $d=0.19\lambda_p$ showing the reflection band of the plasmonic band gap $\omega < \omega_p$ combined with a subwavelength grating broadband reflection feature above $\omega_p$.}
    \label{fig:reflectivity_thickness}
    
\end{figure}

In Fig. \ref{fig:reflectivity_thickness}(a), we plot the reflectivity spectrum of a set of metafilm structures with varying thickness $d$. We note that the band gap is a property of an infinite structure $d \rightarrow \infty$. Also, the surface plasmon resonance is localized to the upper interface between the metafilm and air. Both features are largely independent of the thickness $d$ as long as $d$ is larger than the penetration length, which we observe in Fig. \ref{fig:reflectivity_thickness}(a). As the thickness of the structure drops below $d=0.07/\lambda_p$, approximately the smallest penetration depth observed in Fig. \ref{fig:intro}(e), there is a clear narrowing in the bandwidth of the metafilm's high-reflection band.

In the band structure analysis conducted thus far, we have largely focused on the frequency range where $\omega < \omega_p$, in which the metal has a negative dielectric constant. In the frequency range above $\omega > \omega_p$, the dielectric constant of the metal, as described by the Drude model, becomes positive. The metafilm structure in this case behaves as a dielectric grating which supports many guided resonances \cite{Fan2002,Suh2003}. These guided resonances result in strong reflection at frequencies above $\omega_p$, as shown in Fig. \ref{fig:reflectivity_thickness}(a). In particular, with the right choice of thickness (e.g $d \approx 0.19 \lambda_p$) these guided resonances can provide broad-band reflection \cite{Mateus2004, Magnusson2008, Karagodsky2010}, which in this case extends from $\omega \approx \omega_p$ to $\omega \approx 1.6 \omega_p$, as shown in Fig. \ref{fig:reflectivity_thickness}(b). Thus, one can combine the band gap behavior below $\omega_p$ with the guided resonance behavior above $\omega_p$ to further extend the bandwidth for achieving high reflection.

\section{Extension to Real Materials}

Up to now, we have shown a significantly reduction in the penetration depth of the meta-film structure, which consists of both metal and dielectric, as compared to a metal film structure. The metal in the previous calculation is described using the Drude model, whereas the dielectric is assumed to be lossless and dispersionless. In this section, we show that similar effect can be achieved in realistic material systems. 

\begin{figure}[h]
    \centering
    \includegraphics[width=2.7 in, height = 4 in]{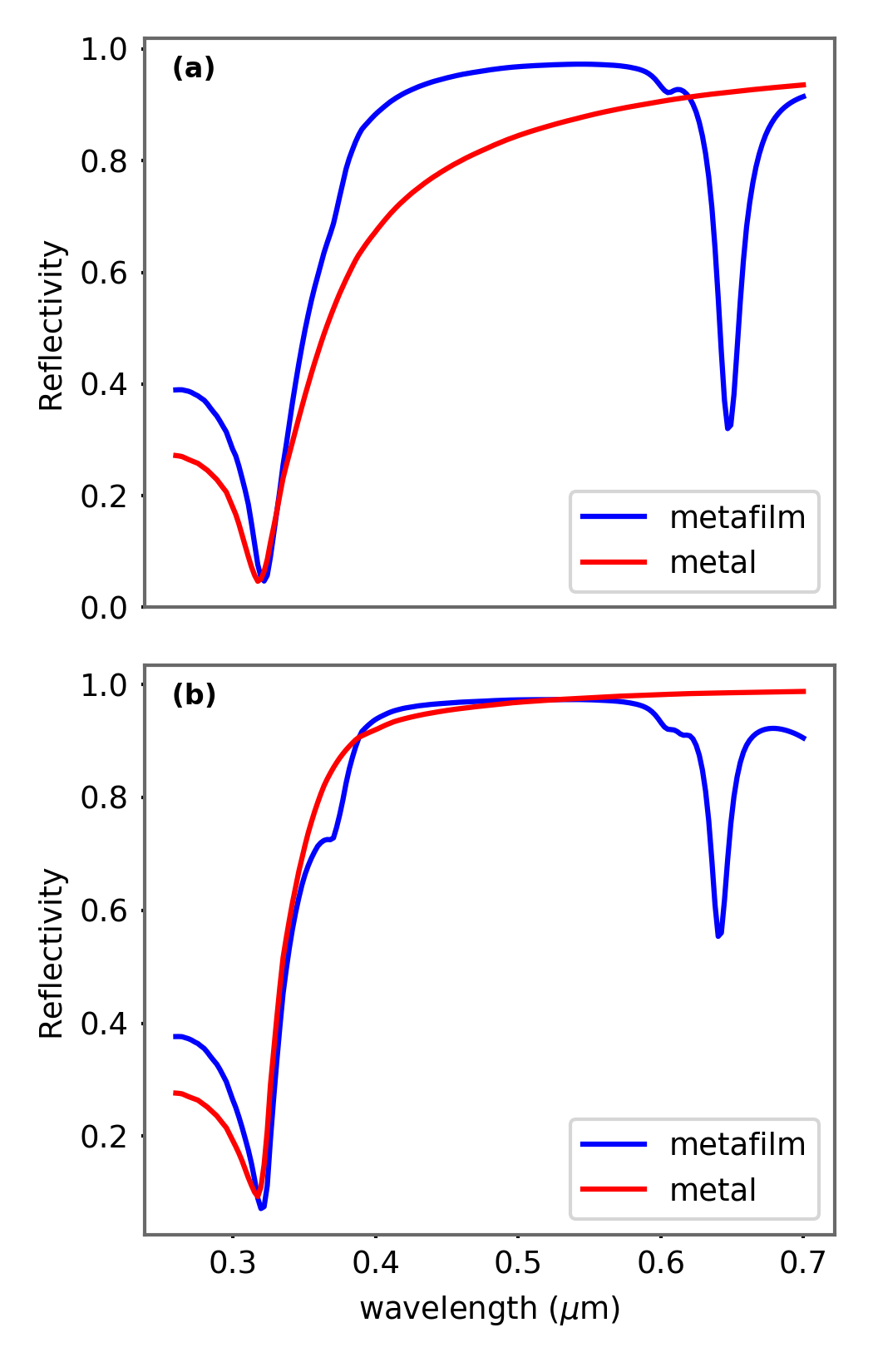}
    \caption{(a) Reflectivity spectrum of a metafilm with $d=30$ nm (blue) compared to that of a uniform metal film of the same thickness (red). The metafilm consists of silicon and silver as described by tabulated dielectric function this wavelength range while the uniform metal film consists of silver with the same material parameters. (b) Reflectivity spectrum of the metafilm with $d=60$ nm with the same material parameters as in (a) compared with the analagous spectrum of a uniform metal film of the same thickness.}
    \label{fig:real_spectra}
\end{figure}

For illustrative purposes, we assume the metal to be silver, and the dielectric to be silicon. We use the experimentally measured permittivity for both materials, fully accounting for absorption \cite{Babar2015, Schinke2015}. We consider a metafilm structure similar to Fig. \ref{fig:intro}(b), with a thickness $d =30$ nm, a lattice constant $a = 50$ nm, and a metal fraction $f = 0.5$. The reflectivity spectrum for such a structure is shown as the blue line in Fig. \ref{fig:real_spectra}(a), and is compared with the reflectivity spectrum of a silver film of equivalent thickness. The reflectivity spectrum of the meta-film structure is qualitatively similar to the system considered earlier using the Drude model. In particular, the reflectivity spectrum shows a band gap behavior in the wavelength range between 350 nm and 600 nm. Over a significant part of this visible wavelength range from 350 nm to 600 nm, which lies within the band gap, the meta-film structure shows significant enhancement of reflection as compared to the uniform metal thinfilm. This enhancement occurs in spite of the significant absorption loss for silicon in this wavelength range, and can be attributed to the reduced penetration depth inside the meta-film as compared to that in a uniform metal structure. 

Since the use of the meta-film structure results in a reduced penetration depth as compared to the uniform metal film, the reflection enhancement in the meta-film structure should be most prominent when the thickness of the films are comparable to the penetration depth, and such enhancement should not be prominent for a much thicker film. We illustrate this effect in Fig. \ref{fig:loss_analysis}(b), where we plot the reflectivity spectra for a metafilm and a uniform metal film that are the same as those in Fig. \ref{fig:loss_analysis}(a), except with a thickness $d = 60$ nm. We see that the reflectivity of the two structures is now comparable, with the remaining contribution being largely accounted for by the absorption of the structure. In principle, the metafilm's absorption within the band gap, can always be reduced below that of the uniform metal film by using a higher index dielectric. This further increases the effective mass of the upper band-edge and further drives down the penetration depth, reducing material absorption and guaranteeing that the reflection of the metafilm at large thicknesses is still superior to that of the uniform metal film.

\section{Application to Hollow Core Waveguides}
As one application of the low penetration depth in the metafilm, we now show it can be used to achieve high packing density of multiple hollow-core waveguides. We note that other works have attempted to use high reflectivity films to make hollow core waveguides \cite{Zhou2009}. However, the design in Ref. \citenum{Zhou2009} experiences decreasing reflectivity for increasing non-normal incidence angles at a single frequency, since in that case the reflectivity arises from guided resonances. The metafilm on the other hand, does not experience this issue since its operating principle is based on a photonic band gap. 


Our demonstration involves two adjacent hollow core waveguides separated by a thin wall of thickness $t= 0.05$ $\mu$m. We use an operating wavelength of $\lambda =0.45$ $\mu$m and a hollow core width of $0.30$ $\mu$m. The metal is silver while the dielectric is that of silicon as used in the previous section. We excite the modes of the waveguide by using a point source located within the upper waveguide (top channel) and assess how much energy leaks into the lower channel (bottom channel) per unit length. In this application, the leakage is proportional to the length scale at which the power in the upper waveguide completely transfers or ``leaks'' to the lower waveguide, and vice versa. Higher leakage would mean the power oscillates between the two waveguides with a high spatial frequency.

\begin{figure*}[htbp]
    \centering
    \includegraphics[]{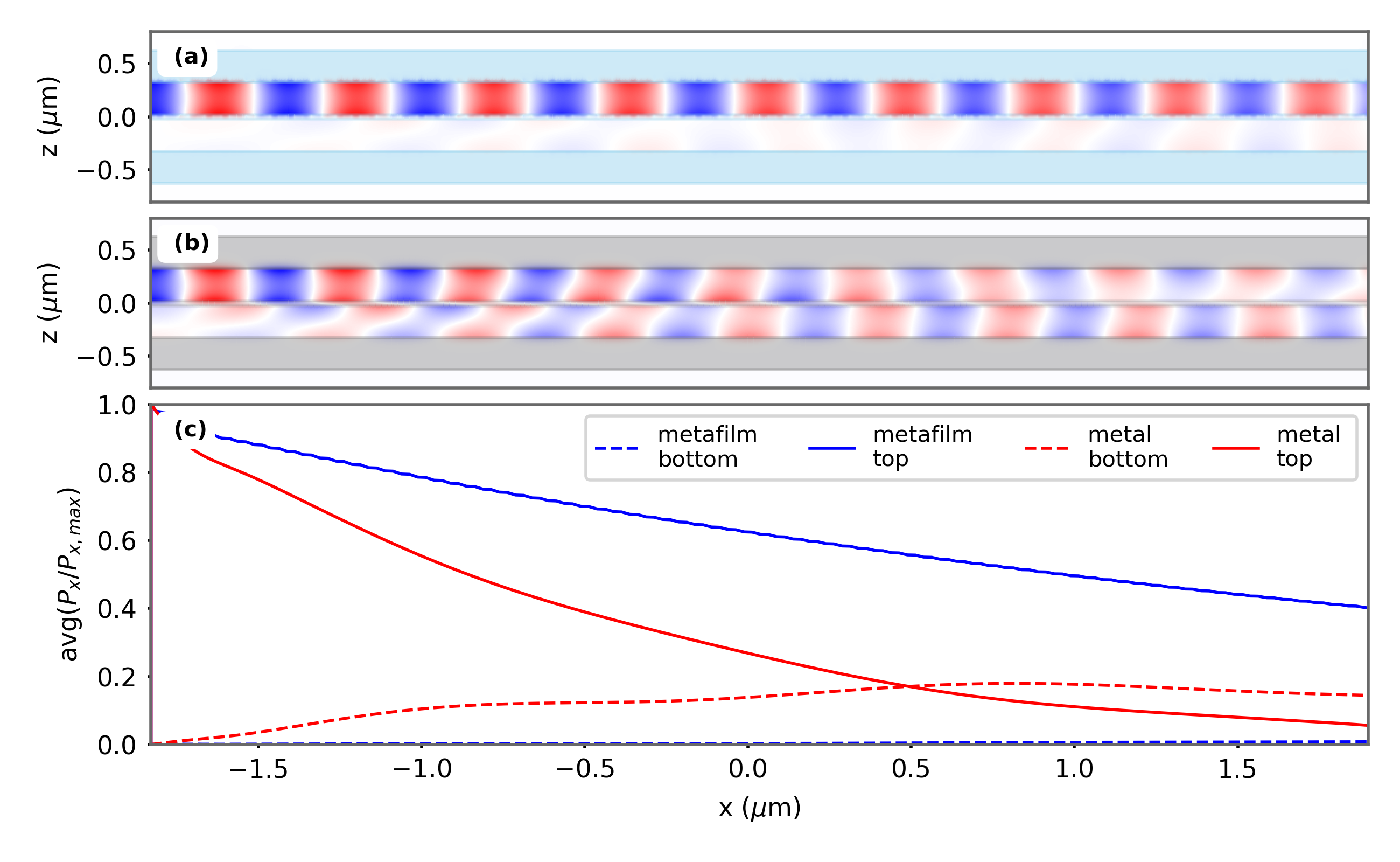}
    \caption{(a) Distribution of Re($H_y{(x,z)}$) for a pair of hollow core waveguides with walls consisting of the metafilm with $\lambda =0.45$ $\mu$m, a central wall thickness of $50$ nm, and an air core width of $0.3$ $\mu$m. We use the same geometric parameters for the metafilm as in the previous section. (b) Distribution of Re($H_y{(x,z)}$) for the same structure but with metal walls. In a we mark the metafilm regions in blue and in b we mark the metal regions in gray. (c) The spatially-averaged power flux in the air region in the $x$-direction for the upper waveguide and lower waveguide in a and b. We truncate the plots for all panels at the source on the left hand side and at the perfectly matched layer boundary on the right hand side.}
    \label{fig:dual_core}
\end{figure*}

From the field distribution shown in Fig. \ref{fig:dual_core}(a), we observe that the mode supported by the metafilm-clad waveguide resembles a TEM$_{00}$ mode of a hollow core waveguide with perfect electric conductor (PEC) walls, in the sense that the field is more uniform inside the air region. This observation is remarkable given that this system operates far from the regime where the metal behaves as a good conductor. In contrast, the field distribution shown in Fig. \ref{fig:dual_core}(b) indicates that the mode supported by the uniform metal waveguide is more plasmonic, due to the higher field concentration at the metal-air interface versus the center of the air core. In other words, it appears more like the surface plasmons at the two metal-air surfaces coupled together. This contrast is consistent with the reduced penetration depth of the metafilm as predicted in our previous analysis.

Moreover, the smaller penetration depth of the metafilm clearly results in suppressed coupling between the two waveguides. A comparison of Fig. \ref{fig:dual_core}(a),(b) indicates that the packed metafilm-air-metafilm waveguides leaks less power into the lower waveguide than the metal-air-metal waveguides. This is confirmed by examining the spatial evolution of the average flux in the $x$-direction shown in Fig. \ref{fig:dual_core}(c). The solid blue line indicates the power in the upper metafilm waveguide while the solid red line indicates the same for the metal waveguide. The dashed blue line indicates the power in the lower channel of the metafilm waveguide, which only experiences a negligible increase after nearly $4$ $\mu$m of propagation. Meanwhile, the dashed red line, indicating the power in the bottom channel of the metal film waveguide, shows much more significant power leakage. In fact, the difference in the power at $x=1.0$ $\mu$m in Fig. \ref{fig:dual_core}(c) in the lower waveguide between the metafilm-air-metafilm waveguide and the metal-air-metal waveguide is more than a factor of 10 even in the presence of considerably higher losses in the metal waveguide. In particular, Fig. \ref{fig:dual_core}(c) also demonstrates that the total propagation loss of the metafilm-air-metafilm waveguide is smaller by slightly less than a factor of two, particularly because the regime of high reflection in the metafilm coincides with the regime where plasmonic surface states in the uniform metal film are evanescent in the $x$-direction.


\section{Conclusion}
In conclusion, we have shown the penetration depth in a metafilm structure, consisting of alternating regions of metal and dielectric, can be significantly reduced as compared to that of a regular metal film. The reduction in the penetration depth arises due to the enhanced effective mass in the photonic band structure of a periodic plasmonic waveguide array. The effect of penetration depth reduction persists with realistic material properties including both dispersion and loss. This effect can be used to enhance the reflectivity of thin film system, and to reduce the cross-talk between densely packed plasmonic waveguides. 


This work was supported by the DOE `Photonics at Thermodynamic Limits' Energy Frontier Research Center (Grant N\textsuperscript{\underline{o}} DE-SC0019140) and by an AFOSR MURI project (Grant N\textsuperscript{\underline{o}}. FA9550-17-1-0002). 

\input{arxiv_submission.bbl}

\end{document}

%% file: arxiv_submission.bbl
\providecommand{\latin}[1]{#1}
\makeatletter
\providecommand{\doi}
  {\begingroup\let\do\@makeother\dospecials
  \catcode`\{=1 \catcode`\}=2 \doi@aux}
\providecommand{\doi@aux}[1]{\endgroup\texttt{#1}}
\makeatother
\providecommand*\mcitethebibliography{\thebibliography}
\csname @ifundefined\endcsname{endmcitethebibliography}
  {\let\endmcitethebibliography\endthebibliography}{}